\documentclass[twocolumn]{aastex631}

\usepackage{graphicx}
\usepackage[varg]{txfonts}
\usepackage{cases}
\usepackage{natbib}
\usepackage{multirow}
\usepackage{longtable}
\usepackage{booktabs}

\shorttitle{Transverse oscillation of prominence and filament induced by an EUV wave from the farside of the Sun}
\shortauthors{Zhang et al.}

\graphicspath{{./}{figures/}}

\begin{document}

\title{Transverse oscillation of prominence and filament induced by an EUV wave from the farside of the Sun}

\correspondingauthor{Yanjie Zhang}
\email{zhangyj@pmo.ac.cn}

\author[0000-0003-1979-9863]{Yanjie Zhang}
\affiliation{Key Laboratory of Dark Matter and Space Astronomy, Purple Mountain Observatory, CAS, Nanjing 210023, China}

\author[0000-0003-4078-2265]{Qingmin Zhang}
\affiliation{Key Laboratory of Dark Matter and Space Astronomy, Purple Mountain Observatory, CAS, Nanjing 210023, China}
\affiliation{Yunnan Key Laboratory of Solar physics and Space Science, Kunming 650216, People's Republic of China}

\author[0000-0002-0786-7307]{De-chao Song}
\affiliation{Key Laboratory of Dark Matter and Space Astronomy, Purple Mountain Observatory, CAS, Nanjing 210023, China}

\author[0000-0002-5898-2284]{Haisheng Ji}
\affiliation{Key Laboratory of Dark Matter and Space Astronomy, Purple Mountain Observatory, CAS, Nanjing 210023, China}

\begin{abstract}
{In this paper, we report our multi-angle observations of the transverse oscillation of a prominence and a filament induced by an EUV wave originating from the farside of the Sun on 2014 September 1.
The prominence oscillation was simultaneously observed by both Atmospheric Imaging Assembly (AIA) onboard the Solar Dynamics Observatory (SDO) spacecraft and Extreme-UltraViolet Imager (EUVI) onboard the Behind Solar Terrestrial Relations Observatory (STEREO) spacecraft.
The speed of the shock travelling in the interplanetary space exceeds that of the EUV wave, and the coronal dimming area experiences minimal growth. 
This indicates that the shock wave is driven by the CME, while the EUV wave freely propagates after the lateral motion of the CME flanks has stopped.
The observed oscillation direction of the prominence, determined through three-dimensional reconstruction, further supports this point.
Moreover, The detailed investigation of the oscillations in the prominence and filament induced by the EUV wave reveals initial amplitudes of 16.08 and 2.15 Mm, periods of 1769 and 1863 s, damping time scales of 2640 and 1259 s, and damping ratios of 1.49 and 0.68, respectively.
The radial component of magnetic field, as derived from the prominence and filament oscillation measurements, was estimated to be 5.4 G and 4.1 G, respectively.
In turn, utilizing the onset times of both the prominence and filament oscillation, the average speeds of the EUV wave are determined to be 498 km s$^{-1}$ and 451 km s$^{-1}$, respectively.
}

\end{abstract}


\section{Introduction} \label{sec:Int}
Solar prominences are often observed as bright elongated emission against the dark background at the solar limb in chromospheric and coronal lines \citep{Tan1995,Mar1998,Lab2010,Che2020}.
They characterise dark features, called filaments, on the disk because they absorb intense solar atmospheric radiation.
Depending on the location where a filament is formed, they can generally be divided into three types: active region filaments (AFs) \citep{Yan2015}, quiescent filament (QFs) \citep{Hei2008} and intermediate filament (IFs).
The prominence and filament are the same entities and we use the term ``prominence'' throughout the paper.

\begin{figure*}
\begin{interactive}{animation}{fig1anim.mp4}
\plotone{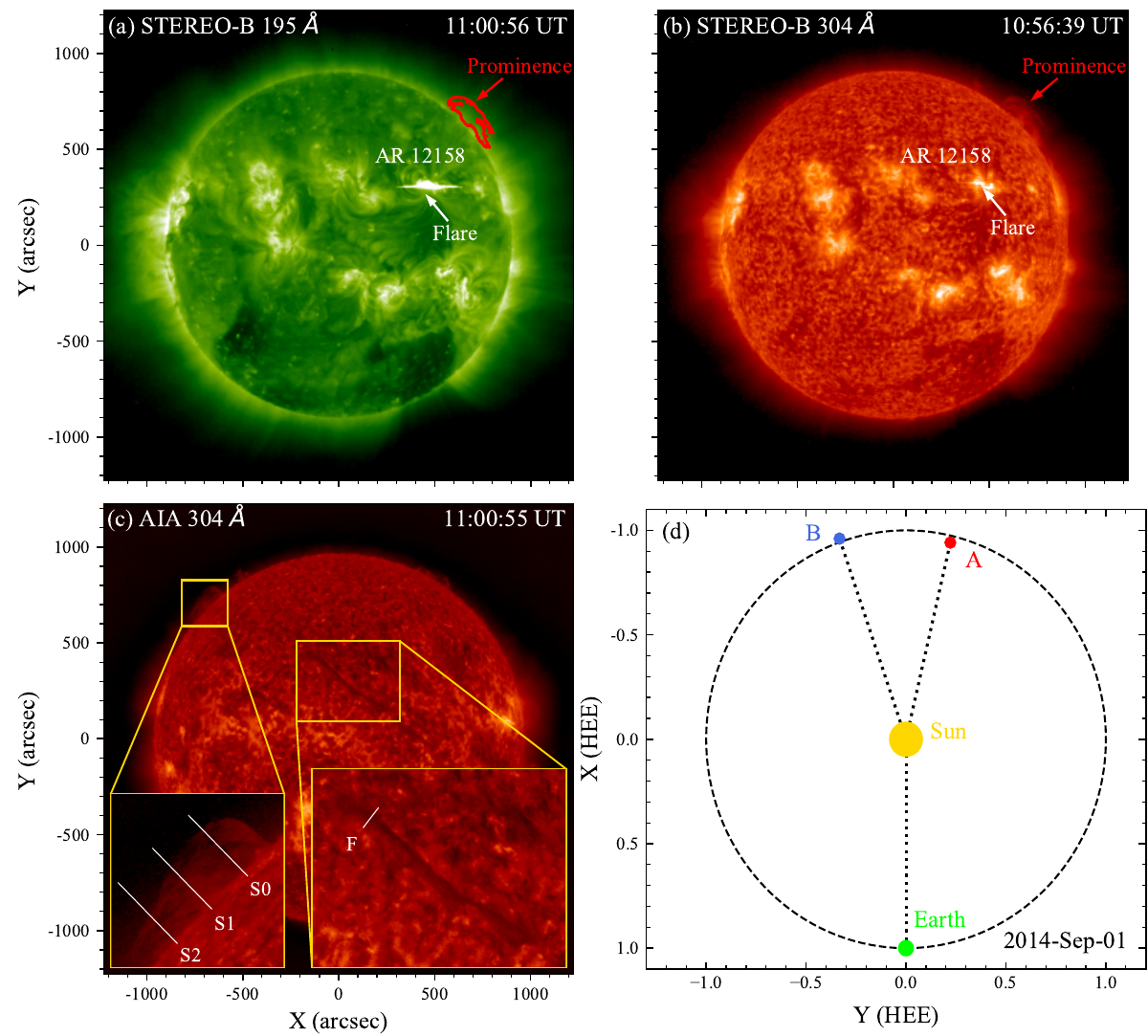}
\end{interactive}
\caption{(a and b) Images of STEREO-B 195 {\AA} at 11:00 UT and 304 {\AA} at 00:56 UT on 2014 September 1, respectively.
The location of AR 12158 is labelled in each panel, as well as the flare and the prominence.
(c) Image of AIA 304 {\AA} at 11:00 UT.
{The subfigures in the panel show an enlarged view of the prominence, identical to that in STEREO-B images, as well as a filament on the disk, respectively.
Three slices, S1, S2 and S3, are utilized to plot the time-distance diagram for the prominence oscillation, while F is for the filament.}
(d) Position of the Earth, Ahead (A), and Behind (B) STEREO spacecraft at 11:17 UT on 2014 September 1.
An animation of the EUVI 195 Å and AIA 304 Å images, available in the online Journal, illustrates the global evolution of the EUV wave from the STEREO-B perspective and provides a close-up view of the prominence oscillation from Earth's perspective.
\label{fig1}}
\end{figure*}

Prominences are rich in dynamics, in which their large-amplitude oscillations induced by EUV waves in association with remote flaring activities has been a hot topic of research \citep{Wil2009,Gal2011,Pat2012,Liu2014,War2015,Che2016}.
When the EUV wave was first discovered, it was also referred to as the EIT wave, named after the observing telescope Extreme-ultraviolet Imaging Telescope \citep[EIT;][]{Del1995} onboard the Solar and Heliospheric Observatory (SOHO).
They are observed as bright wavelike fronts propagating across most of the solar disk in SOHO/EIT running-difference images if the coronal magnetic structure is simple \citep{Tho1998}.
They occur following flare/coronal mass ejection (CME), accompanied by extending dimming region behind them \citep{Tho2000}.
Initially, EUV waves were explained naturally to be the coronal counterparts of chromospheric Moreton waves, that is coronal fast-mode (shock) waves \citep{Mor1960,Uch1968,Uch1974,Tho2000,Wan2000,Wu2001,Ofm2002}.
However, many EUV waves exhibit properties distinct from those of Moreton waves \citep{Eto2002,Zha2011}, giving rise to various conjectures about their physical nature \citep{War2001,Vrs2002,Tri2007,War2007,Wil2009,Gal2011}.
{Currently, a small fraction of events has demonstrated the presence of two components in EUV waves, one fast and one slow\citep{Che2011,She2012b,She2014}, while the majority of events have not exhibited this feature.}

Large amplitude prominence oscillations caused by EUV waves are often observed \citep{Eto2002,Oka2004}. 
Specifically, prominence oscillations can be further classified into longitudinal and transverse oscillations.
{Longitudinal oscillations refer to the oscillatory motion of prominence material along the guiding toroidal magnetic field, with most events not exceeding 20$^{\circ}$ \citep{Jin2003,Li2012,Lun2012,Zha2012,Lun2014,Dai2021}.
However, recent studies have also reported longitudinal oscillations exceeding 20$^{\circ}$ \citep{She2019,Tan2023}.}
In contrast, transverse oscillations occur in the direction perpendicular to the spine \citep{Hyd1966,Ram1966,Kle1969,She2014}.
Recently, the physical nature of prominence oscillations, such as triggering mechanisms, restoring forces, and damping mechanisms, has been effectively investigated using state-of-the-art magnetohydrodynamics (MHD) numerical simulations \citep{Ter2013,Ter2015,Zha2013,Lun2016,Zho2017,Zho2018,Dev2022,Lia2023}.

Until now, oscillations in prominence {and filament} induced by {EUV waves} originating from the farside of the Sun have not been comprehensively investigated.
In this study, we analyzed the prominence {and filament} oscillation associated with a farside EUV wave on 2014 September 1. 
The prominence was simultaneously observed by the Behind Solar Terrestrial Relations Observatory \citep[STEREO-B, hereafter STB;][]{Kai2008} spacecraft and the Solar Dynamics Observatory \citep[SDO;][]{Pes2012}, {while the flare was only observed by STB and the filament was exclusively observed by SDO.}
{We found that the EUV wave freely propagated once the lateral motion of the CME flanks stopped. 
In contrast, the shock wave in the interplanetary space is all driven by the CME, and its speed exceeds that of the EUV wave.
The speeds of the EUV wave reaching the prominence and filament were calculated to be 498 km s$^{-1}$ and 451 km s$^{-1}$, respectively.}
Furthermore, the unprecedented time and spatial resolution of multi-wavelength EUV data recorded with SDO/AIA enable us to accurately determine the parameter of the prominence and filament oscillation, thereby the surrounding magnetic field strength can be estimated.

Data analysis is described in detail in Section~\ref{sec:dat}. 
Results are shown in Section~\ref{sec:Res}.
The discussion and summary are given in Section~\ref{sec:Dis} and Section~\ref{sec:Sum}.

\begin{figure}
\plotone{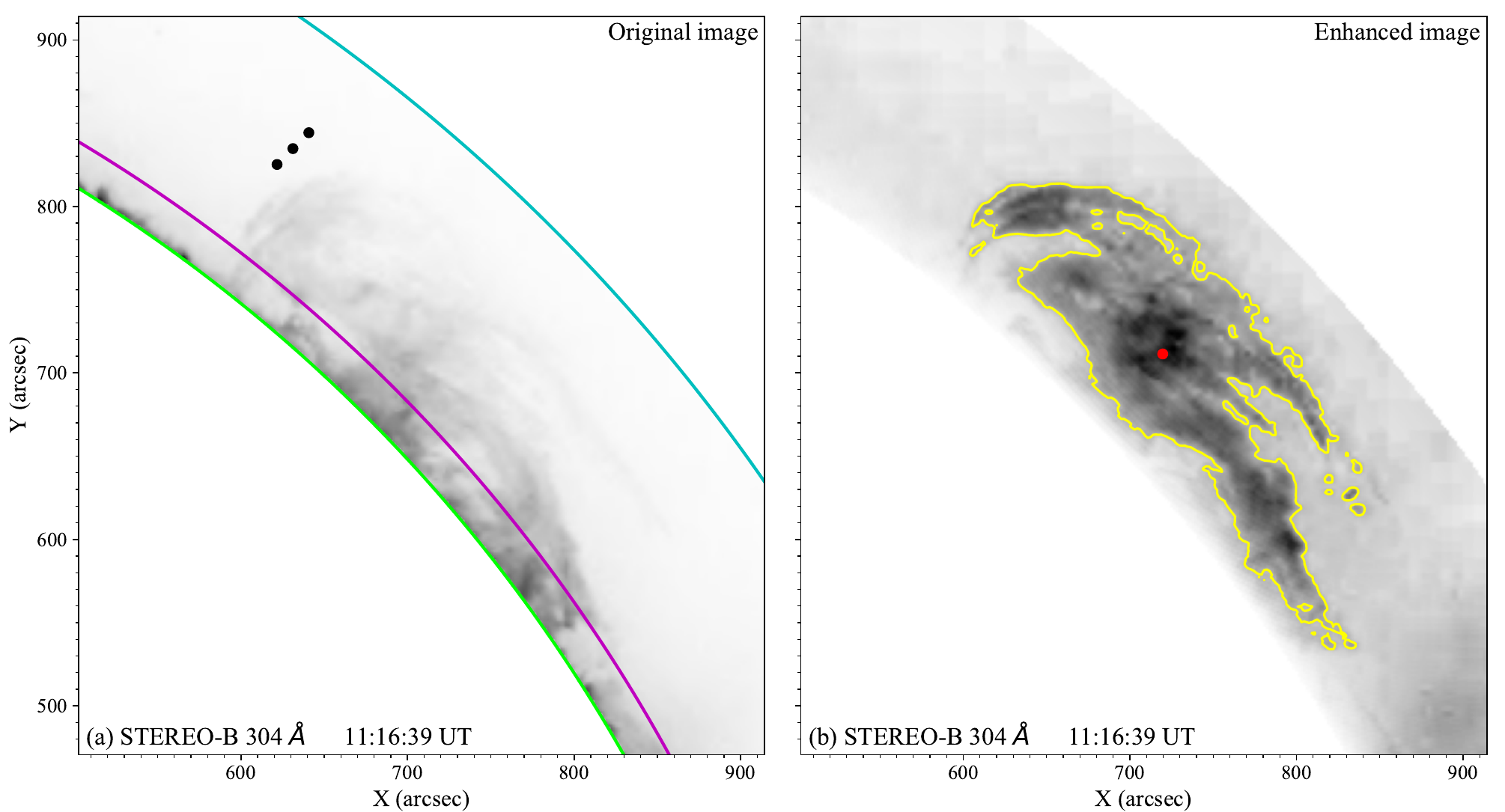}
\caption{(a) Original image of the prominence from EUVI 304 {\AA} at 11:16 UT on 2014 September 1. 
The colored lines are employed to enhance the image in the radial-filter technique, as described in Section~\ref{sec:dat}.
(b) The enhanced image after using the radial-filter technique.
The yellow line represents the contour of intensity threshold set at 2.8 times the average pixel intensity of the enhanced image, and the plus symbol represent the centroid of the prominence by using the contour.
\label{fig2}}
\end{figure}

\begin{figure*}[ht!]
\begin{interactive}{animation}{fig3anim.mp4}
\plotone{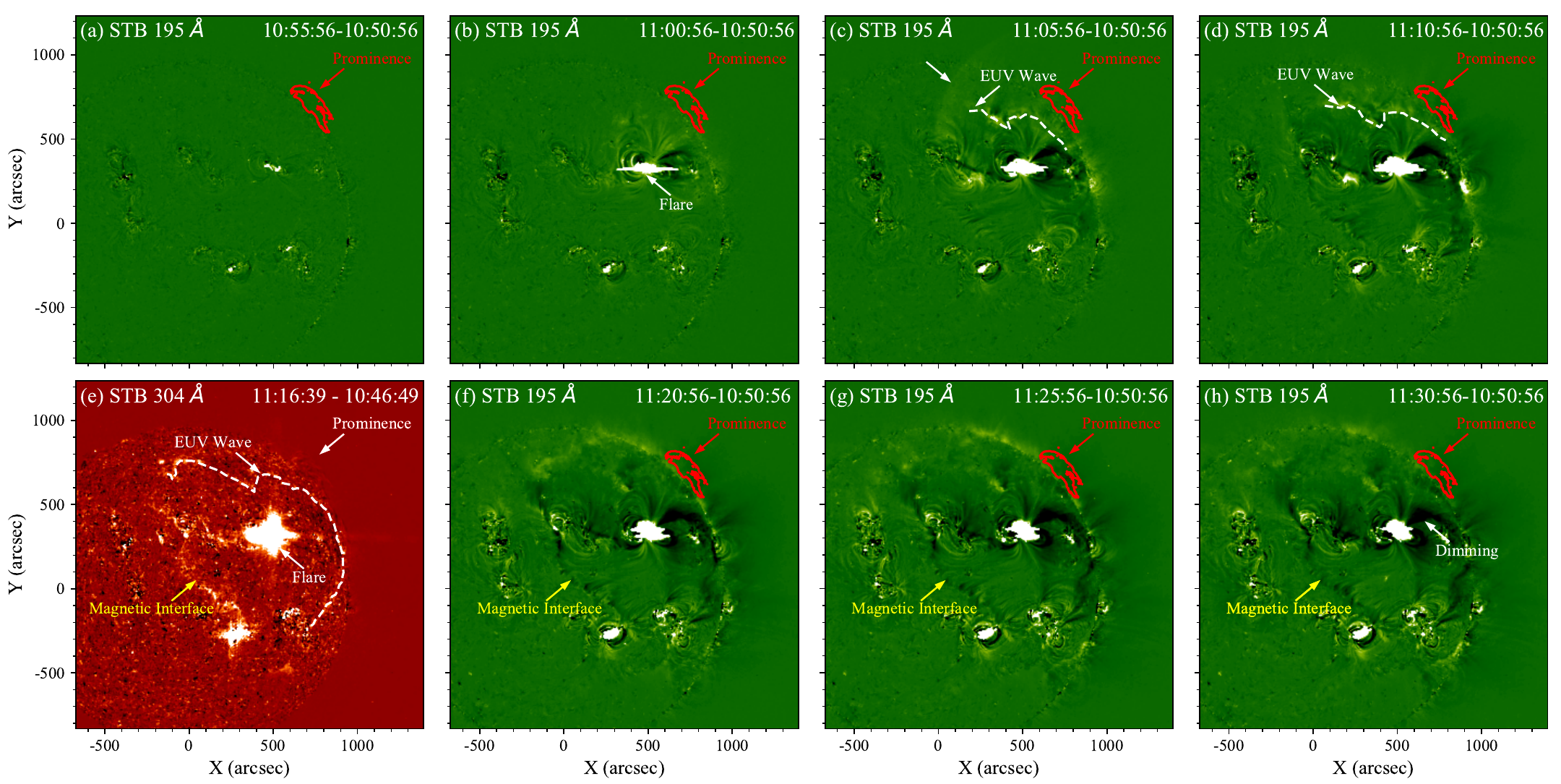}
\end{interactive}
\caption{Detailed feature of atmospheric disturbances in the EUVI 195 {\AA} images subtracted with the image at 10:50:56 UT, except for panel (e) in the EUVI 304 {\AA} image subtracted with the image at 10:46:49 UT. 
The dashed lines depict the wavefront in panel (c), (d) and (e).
An expanding dome is observed in panel (c) and is indicated by an arrow.
Prominence is outlined in each image using the radial-filter technique described in Figure~\ref{fig2}.
There is also a stationary front to the southeast of the disk as the wave front swept by, indicated by a yellow arrow.
An animation of the EUVI 304 Å and 195 Å subtracted images is accessible in the online journal, unveiling additional details in the evolutionary process of the EUV wave.
\label{fig3}}
\end{figure*}

\section{Data Analysis} \label{sec:dat}
The two panels in the first row of Figure~\ref{fig1} show the 195 and 304 {\AA} images of Extreme-UltraViolet Imager \citep[EUVI;][]{Wue2004} of the Sun Earth Connection Coronal and Heliospheric Investigation \citep[SECCHI;][]{How2008} on board the STB on 2014 September 1, where AR 12158 is labelled.
EUVI takes full-disk images out to 1.7\,$R_{\odot}$ with a spatial resolution of $3\farcs2$ in 171, 195, 284, and 304 {\AA}. 
The 195 and 304 {\AA} images have time cadences of 5 and 10 minutes, respectively.
The flare with obvious intensity enhancement is marked with an arrow.
We have no way of knowing the magnitude of the flare from Soft X-ray (SXR) light curves in this period recorded by the Geostationary Operational Environmental Satellite (GOES) spacecraft, because it occurred on the farside of the Sun as shown by panel (d).

The red contour of the prominence in panel (a) is derived from panel (b).
In our case, the intensity at the top of the prominence is lower than the background intensity close to the solar surface in the EUVI 304 {\AA}.
Therefore, it is not feasible to obtain the profile of the prominence by simply selecting an intensity threshold and plotting a contour.
Instead, the radial-filter technique is applied to enhance the emission of the prominence, as demonstrated by Figure~\ref{fig2}.
Specifically, in the pixel coordinates, a series of concentric circles are selected and labelled 1 to 100 from lowest (close to the solar surface) to highest (close to the top of the prominence) as shown by the colored lines in Figure~\ref{fig2}(a).
The intensity values of the EUVI 304 {\AA} in each concentric circle are then multiplied by the number of that concentric circle.
In this way, the intensity at the top of the prominence has been enhanced while that of the background close to the solar surface attenuated, so the contour of the prominence can be well drawn as depicted by Figure~\ref{fig2}(b).

The prominence has also been recorded by the Atmospheric Imaging Assembly \citep[AIA;][]{Lem2012} on board the SDO.
AIA took full-disk images in seven EUV (94, 131, 171, 193, 211, 304, and 335 {\AA}) and two UV (1600 and 1700 {\AA}) wavelengths.
The AIA level\_1 data with a time cadence of 12 s and a spatial resolution of 1$\farcs$2 were calibrated using the standard program \texttt{aia\_prep.pro} in the Solar Software (SSW).
The related CME was simultaneously observed by the Large Angle and Spectrometric Coronagraph \citep[LASCO;][]{Bru1995} on board the SOHO spacecraft and the COR1 white-light (WL) coronagraph on board the STB. 
The separation angle between the STB and Earth was $\sim$161$^{\circ}$ as shown in Figure~\ref{fig1}(d). 
The radio dynamic spectra associated with the flare and CME-driven shock was obtained from the BLENSW ground-based station belonging to the e-Callisto\footnote{http://www.e-callisto.org} network.

\begin{figure*}[ht!]
\plotone{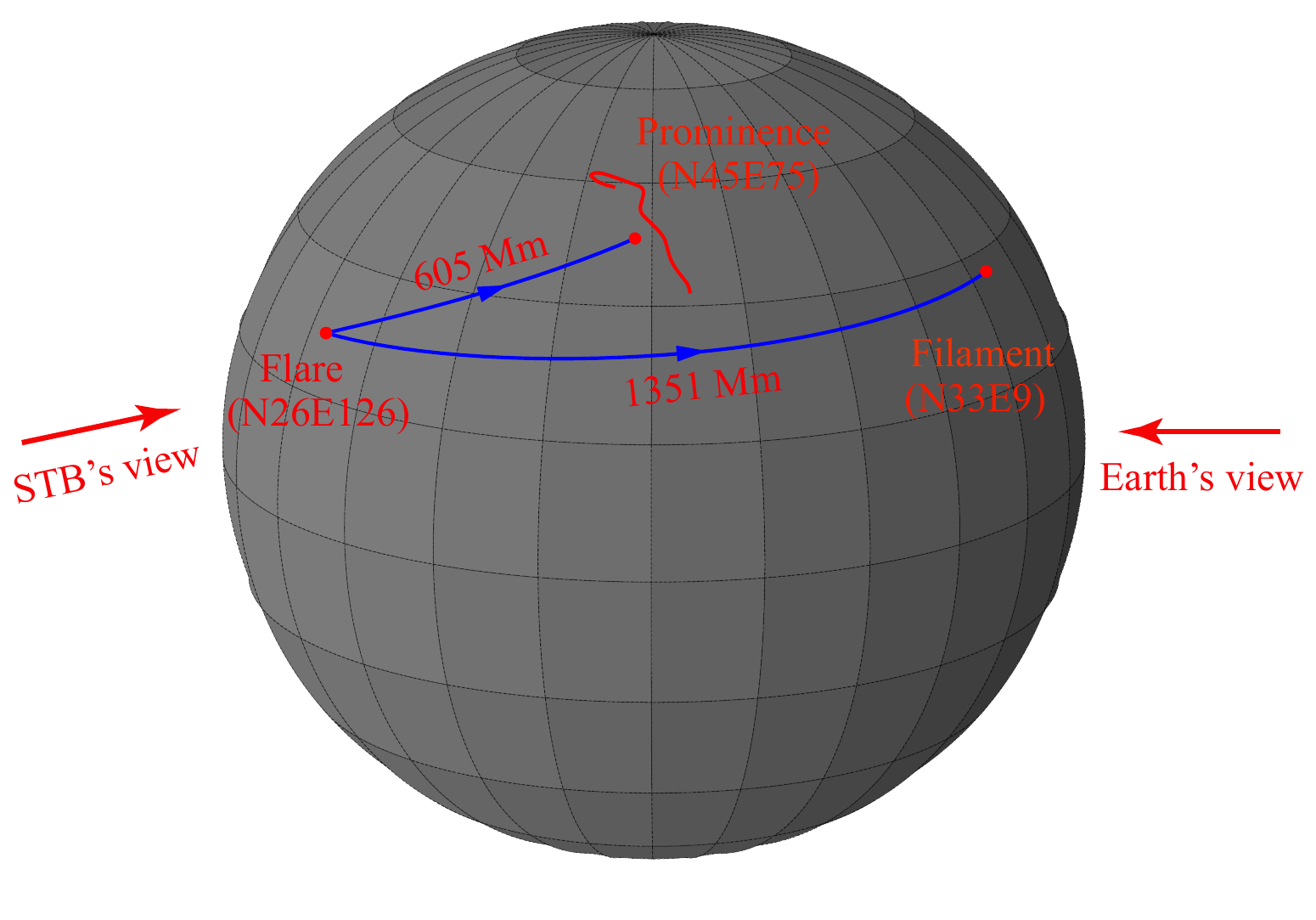}
\caption{{The position of the flare, prominence and filament in the coordinate system from the earth-based vantage point, respectively. The red curve is the result of the 3D reconstruction of the prominence using the AIA and EUVI 304 {\AA} image pair at 11:16 UT.}
\label{fig4}}
\end{figure*}

\section{Results} \label{sec:Res}
\begin{figure*}[ht!]
\plotone{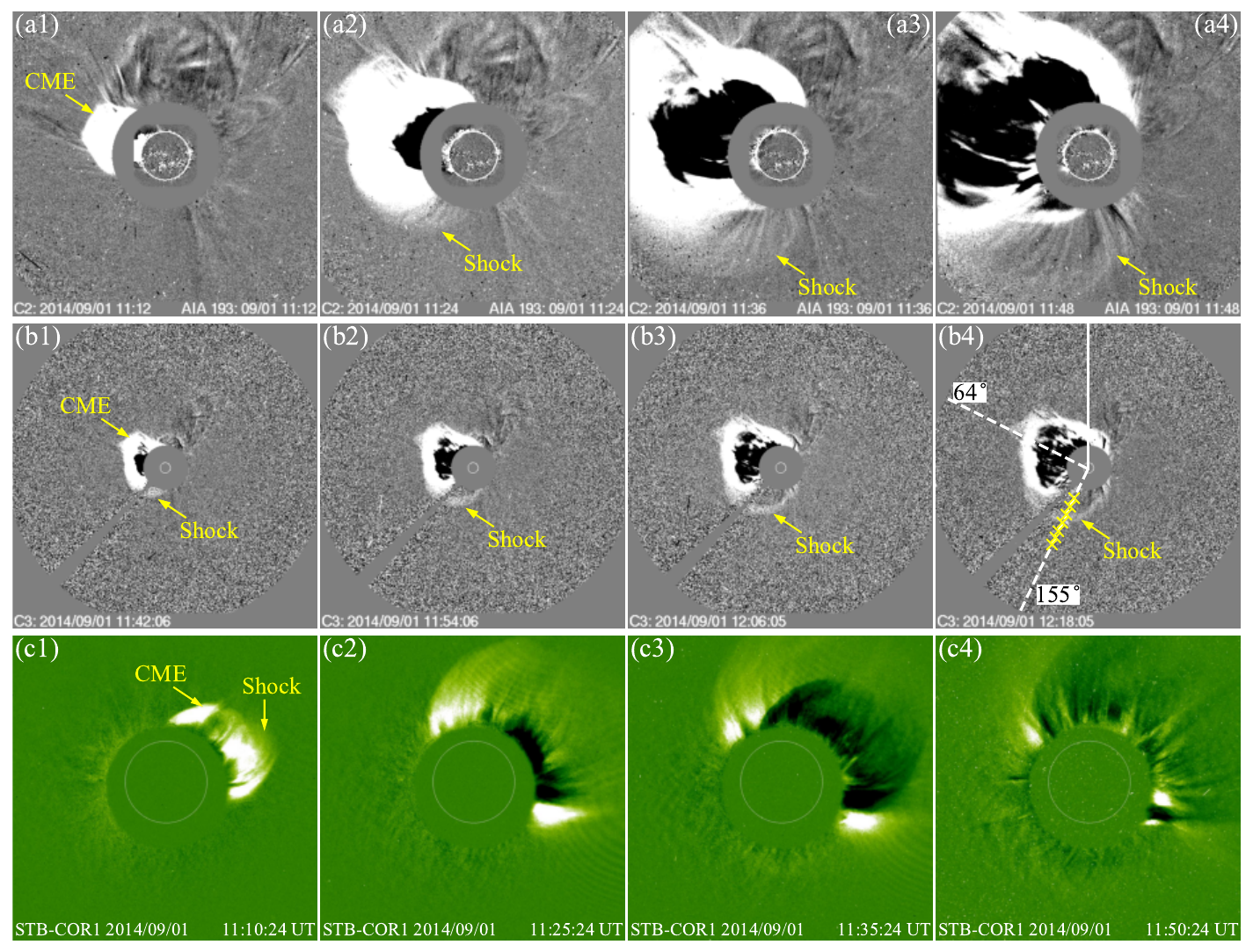}
\caption{LASCO/C2 (top), LASCO/C3 (middle) and STB/COR1 (bottom) difference images showing the halo CME associated with the flare that occurred on the farside of the Sun on 2014 September 1. 
Yellow arrows point to the shock associated with the CME.
Panel (b2) presents the two tracking angles: 64 $^{\circ}$ for the CME in Figure~\ref{fig6}(a) and 155 $^{\circ}$ for the shock, both measured from north (the vertical line).
\label{fig5}}
\end{figure*}

The NOAA active region (AR) 12158 is energetic as it produced many flares during 2014 September 8 to 18 on the visible disk.
The flare we studied also originated from this AR on 2014 September 1, but on the farside of the Sun as shown in Figure~\ref{fig3}.
It should also be a high-energy one, because the flare was accompanied by a halo CME shown in Figure~\ref{fig5}.
In addition, the images of EUVI 195 and 304 {\AA} during flaring time suffered from strong snow-like interference (see the online animation in Figure~\ref{fig1}), probably caused by high-energy particles that hit the image sensor at that time.

Successive base-difference images of EUVI are drawn in Figure~\ref{fig3}. 
All of these images were subtracted from the base image at about 10:50 UT, showing the detailed features of coronal disturbances.
The EUV wave (visible increase front) was first observed at 11:05 UT in panel (c) and then propagated nearly half of the disk along all directions. 
However, it did not show an isotropic feature. 
Specifically, the front is mainly visible to the north of the flare, and a small indentation appeared in its propagation path, which may have been caused by a small AR in that area (see the online animation in Figure~\ref{fig3}).
In addition, a stationary front appeared in the southeast of the flare, where a magnetic interface structure existed before the eruption.
Therefore, this stationary brightening is a result of the propagating disturbance acting on the structure.
However, it can still be observed that a subtle front was passing through the interface in the difference movie (see the online animation in Figure~\ref{fig3}), which demonstrates the nature of the wave.

The dimming appeared near the AR at the same time, while the EUV wave started outside the dimming area, similarly as reported by \citet{Eto2002,Tho1999}.
{It can be observed that the dimming exhibited limited expansion compared to the wave front.
This observation suggests the scenario that the coronal dimming region maps the CME footprint on the solar lower layer, while the EUV wave is a wave driven by the associated CME \citep{Pat2009,Muh2010,Tem2011,She2012b}.}
We can also identify an expanding dome in panel (c), similar to that reported by \citet{Asai2012}.
{The dome is recorded at a only single moment, indicating that its speed must exceed that of the wave front.
This reveals that the dome was propelled by the CME, whereas in the lateral direction, the wave is freely propagating as soon as the lateral expansion of the CME flanks has stopped \citep{Ver2010}.}

{Significant propagating disturbance was also observed in the EUVI 304 {\AA} image in panel (e) and the online animation (Figure~\ref{fig3}).
Considering the similar kinematics to that in the 195 {\AA} images, we propose that the observed wave signature in the 304 {\AA} passband is mainly contributed by the coronal Si \uppercase\expandafter{\romannumeral11} line rather than the chromospheric He \uppercase\expandafter{\romannumeral2} lines \citep{She2012}.}

{The EUV wave induced oscillations in both a prominence and a filament. 
Due to the limited time resolution of the EUVI data and indistinct EUV Wave signals in AIA Data, it is inappropriate to calculate the speed by stacking the time-distance diagram.
Instead, the average velocity of the wave can be estimated by determining the position of the flare, the prominence and the filament.
The position of the filament (the position of the slice F, see Figure~\ref{fig1}(c) ) in the earth-based point is easily discerned (N33E9).
Considering the angle between STB and Earth with respect to the Sun, the location of the flare from the Earth's perspective were calculated as (N26E126) (ignoring the solar axial tilt). 
Applying the three-dimensional (3D) reconstruction program \texttt{scc\_measure.pro} in the SSW software, the center coordinate of the prominence is obtained as (N45E75), as shown in Figure~\ref{fig4}.
Assuming a propagation altitude of 1.1$R_\odot$(the radius of the Sun, about 695500 km) for the EUV wave, the coefficient is sourced from the height of the prominence in the 3D reconstruction.
Therefore, the distances from the flare to the prominence and filament can be determined as 605 Mm and 1351 Mm, respectively.
}

{The initiation time of the flare recorded by EUVI was 10:55:56 UT (see Figure~\ref{fig3}), while the onset times of the prominence oscillation and filament oscillation observed by AIA were 11:16:09 UT and 11:45:57 UT, respectively (detailed discussion to ensue).
Consequently, the average speeds of the EUV wave reaching the prominence and filament can be determined as 498 km s$^{-1}$ and 451 km s$^{-1}$, respectively, with no significant velocity attenuation (assuming the EUV wave originated from the flare site, although their actual source might be at some distance away from the flare center, such as the edge of the dimming region \citep{Tho1999,Eto2002}, this seemingly has limited influence for the calculations).}

\begin{figure*}[ht!]
\plotone{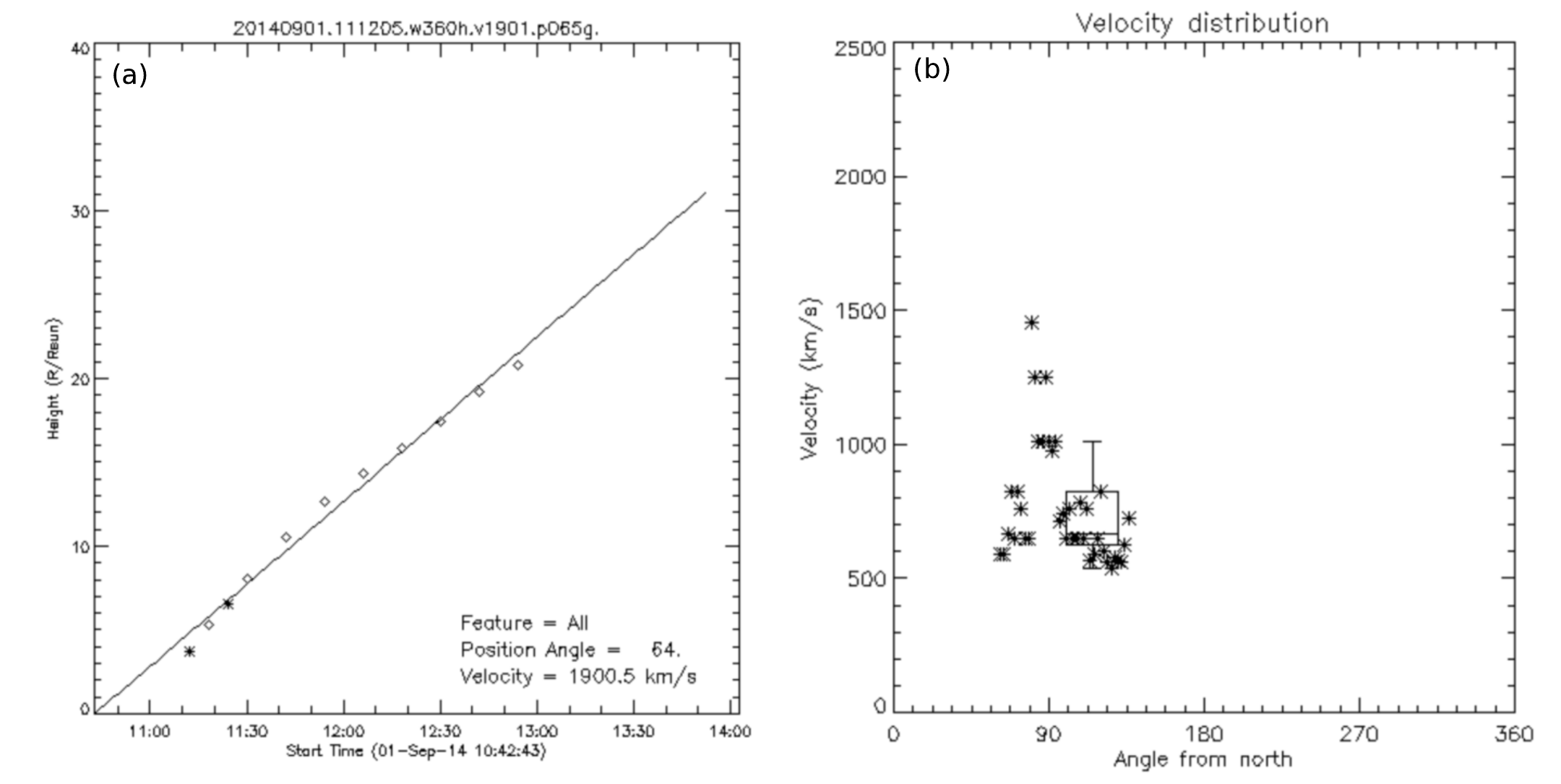}
\caption{(a) The time-distance diagram of the CME generated by the flare, as recorded by the CDAW catalog, with a tracking angle of 64 $^{\circ}$ (as shown in Figure~\ref{fig5}(b4)). (b) The velocity distribution of the CME tracked at different angles by the CACTus website.
\label{fig6}}
\end{figure*}

A halo CME was also produced by the flare as shown in Figure~\ref{fig5}, which was first observed by STB/COR1 at 11:05:48 UT.
It should be noted that in this moment, the prominence had not yet exhibited oscillatory motion.
This indicates that the disturbance propagated faster in the direction of interplanetary space than along the solar surface.
STB/COR1, LASCO/C2 and C3 all observed the shock associated with the CME, as indicated by the yellow arrows in Figure~\ref{fig5}.
The speed of CME is 1901 km s$^{-1}$ recorded by the CDAW catalog $\footnote{http://cdaw.gsfc.nasa.gov/CME\_list}$ where CMEs are recognized manually, 650 km s$^{-1}$ the CACTus website\footnote{http://sidc.oma.be/cactus/scan} where CMEs are identified automatically \citep{Yas2008}.
Figure~\ref{fig6} illustrates the calculation employed by the two websites for the CME velocity determination.
It can be observed that the CDAW catalog involves tracking the fastest direction of CME propagation, whereas the CACTus website averages the velocities across different propagating directions of the CME.
Simultaneously, we also tracked the propagation of the shock (the yellow cross symbols in Figure~\ref{fig5}(b4), representing wavefronts in different images), which exhibited a linear velocity of approximately 814 km s$^{-1}$.

\begin{figure}[ht!]
\includegraphics[width=0.47\textwidth]{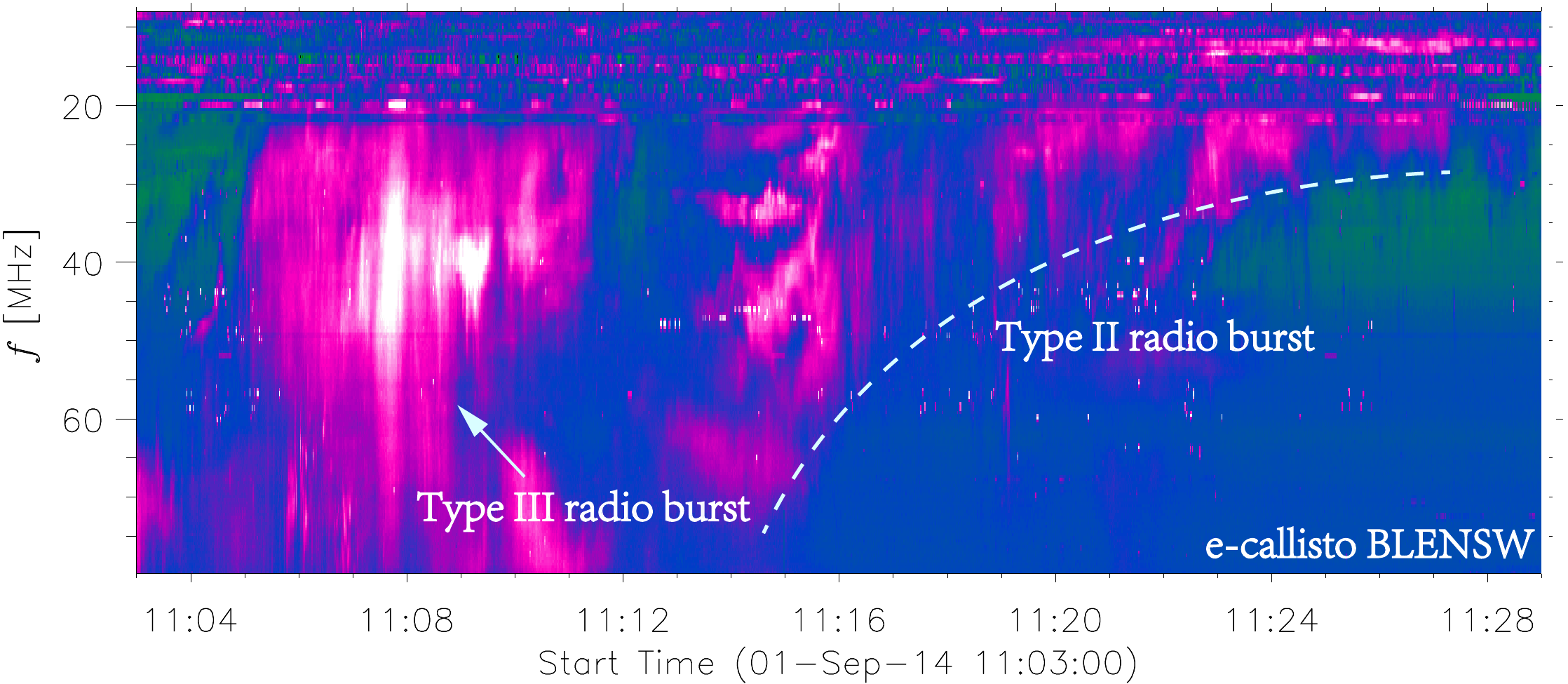}
\caption{Radio dynamic spectra recorded by the e-Callisto/BLENSW station in the frequency of 10 $\sim$ 80 MHz.
The type III radio burst occurred from 11:05 UT to 11:10 UT, while the type II radio burst occurred from 11:14 UT to 11:27 UT.
\label{fig7}}
\end{figure}

\begin{figure}[ht!]
\includegraphics[width=0.47\textwidth]{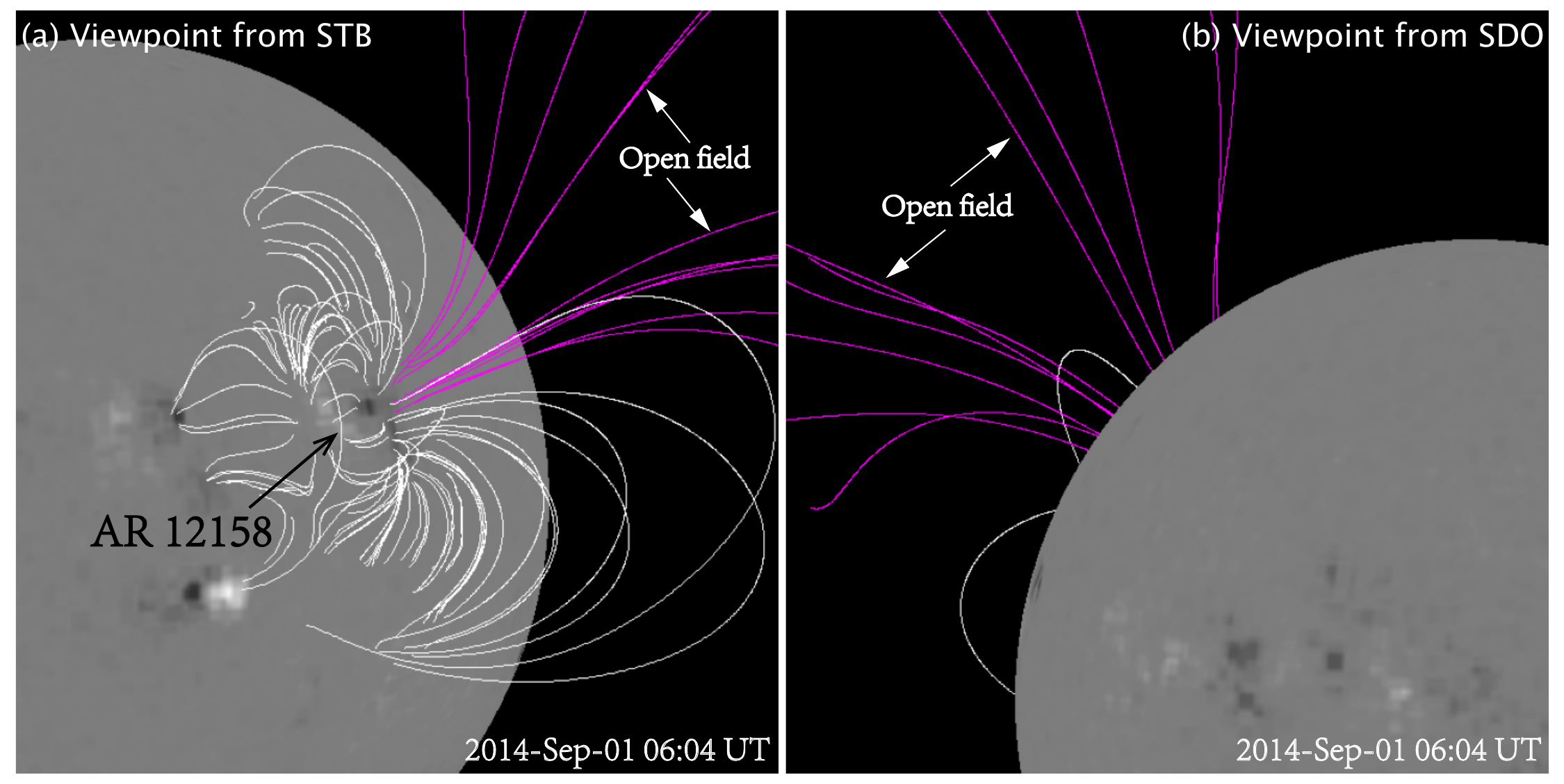}
\caption{PFSS extrapolations from the STB (a) and AIA (b) perspectives at 06:04 UT on 2014 September 1, displays white lines indicating the closed magnetic field lines, and purple lines representing the open magnetic field lines.
The AR where the flare occurred is marked in the figure.
\label{fig8}}
\end{figure}

A flare-related type III and CME-related type II radio burst was detected by the e-Callisto/BLENSW station in the frequency of 10 $\sim$ 80 MHz, which is displayed in Figure~\ref{fig7}.
The flare was estimated to occur in 10:55 UT as shown in Figure~\ref{fig3}. 
Subsequently, the time interval for the Type III radio burst was recorded between 11:05 UT and 11:10 UT, which are thought to be created by plasma emissions of flare-accelerated non-thermal electron beams propagating outward along open field \citep{Kru2011,Mas2013,Zha2015,Wyp2018}.
The existence of open magnetic field lines is confirmed by the Potential Field Source Surface (PFSS) extrapolation, as illustrated by the purple lines in Figure~\ref{fig8}.
The type II radio burst occurred immediately after the type III burst, lasting from 11:14 UT to 11:27 UT, which are associated with the shock waves driven by the fast CME \citep{On2009,Zu2018,Man2019}.

\begin{figure}[ht!]
\includegraphics[width=0.47\textwidth]{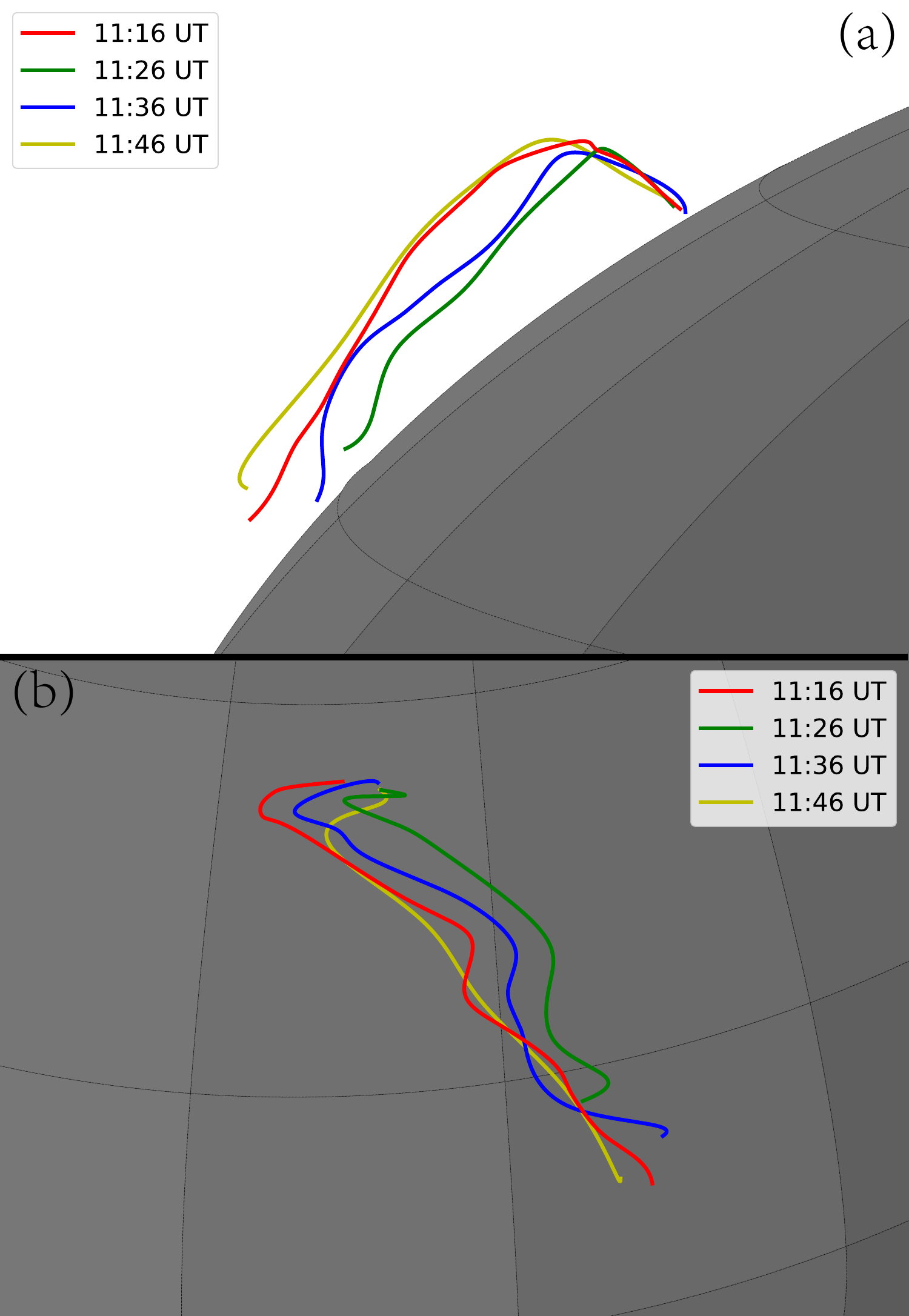}
\caption{{3D reconstruction of the prominence from two distinct vantage points (along the solar limb in panel (a) and perpendicular to the solar surface in panel (b)).}
\label{fig9}}
\end{figure}

{The eruptive event on the solar backside resulted in the oscillation of a prominence and a filament on the solar disk, sequentially. 
As the prominence was observed simultaneously by both AIA and EUVI, a 3D reconstruction was performed, as illustrated in Figure~\ref{fig4}.
Specifically, the oscillation direction can be also determined through the reconstruction, as depicted in Figure~\ref{fig9}. 
During the period from 11:16 UT (red segment) to 11:26 UT (green segment), a distinctive pattern emerges in the prominence's motion, featuring a simultaneous downward vertical motion (panel (a)) and a rightward horizontal motion (panel (b)). 
Subsequently, from 11:36 UT (blue segment) to 11:46 UT (yellow segment), the prominence underwent simultaneous upward vertical motion (panel (a)) and leftward horizontal motion (panel (b)).
It indicates a transverse oscillation.
Moreover, the direction of the oscillation can be calculated as an angle of $\sim$63$^\circ$ to the local photosphere plane.}

\begin{figure*}[ht!]
\plotone{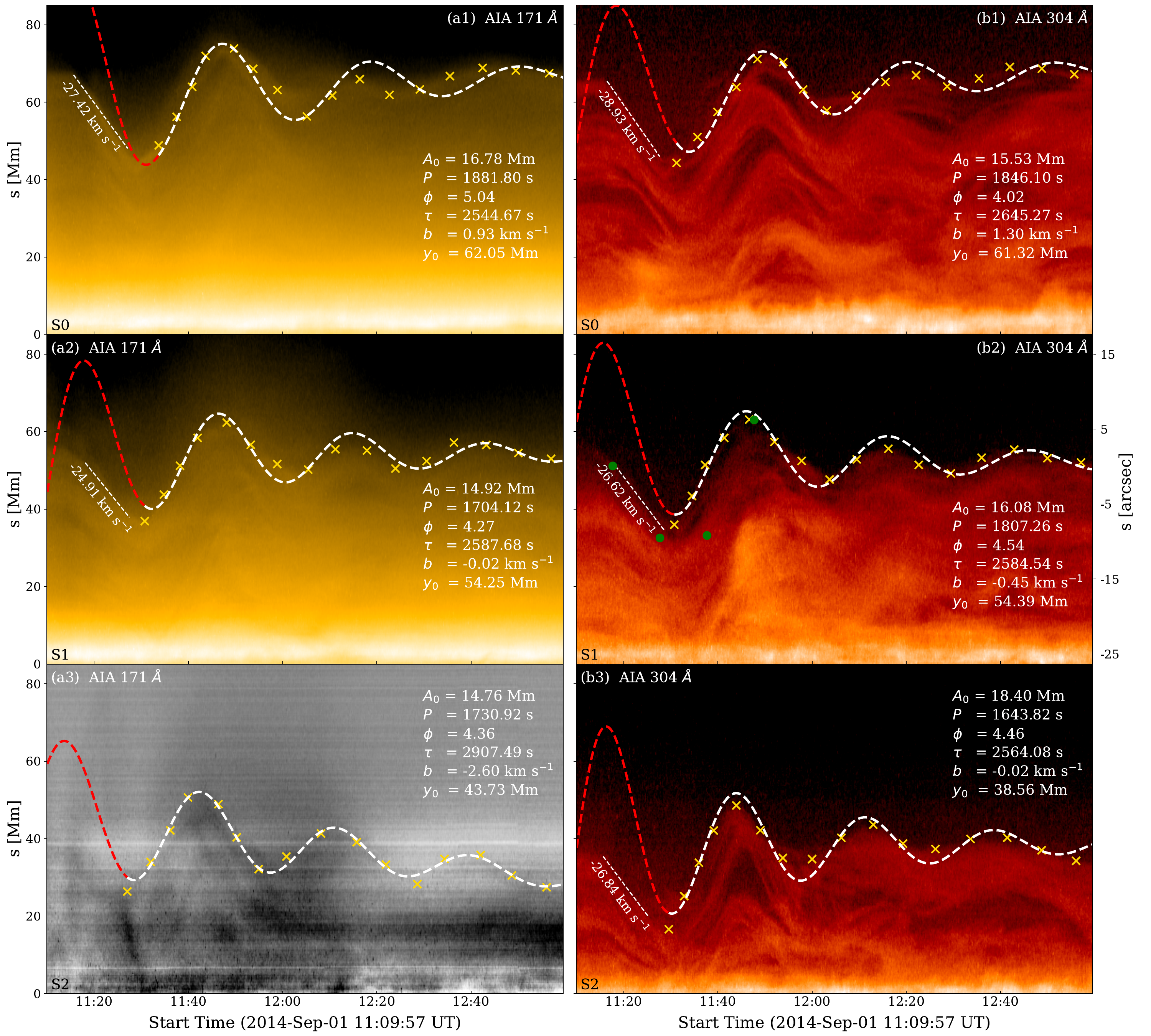}
\caption{Time-distance diagrams of slit S0, S1 and S2 in Figure~\ref{fig1}, showing the prominence oscillation. The dashed lines are fittings of the prominence oscillation by using Equation~\ref{eqn-1}, with the white color indicating a valid fit, while the red color indicates an invalid fit. The four green points in panel (b2) represent the displacement of the prominence centroid in EUVI 304 {\AA} by using the radial-filter technique in Section~\ref{sec:dat}.
\label{fig10}}
\end{figure*}

\begin{figure}
\includegraphics[width=0.49\textwidth]{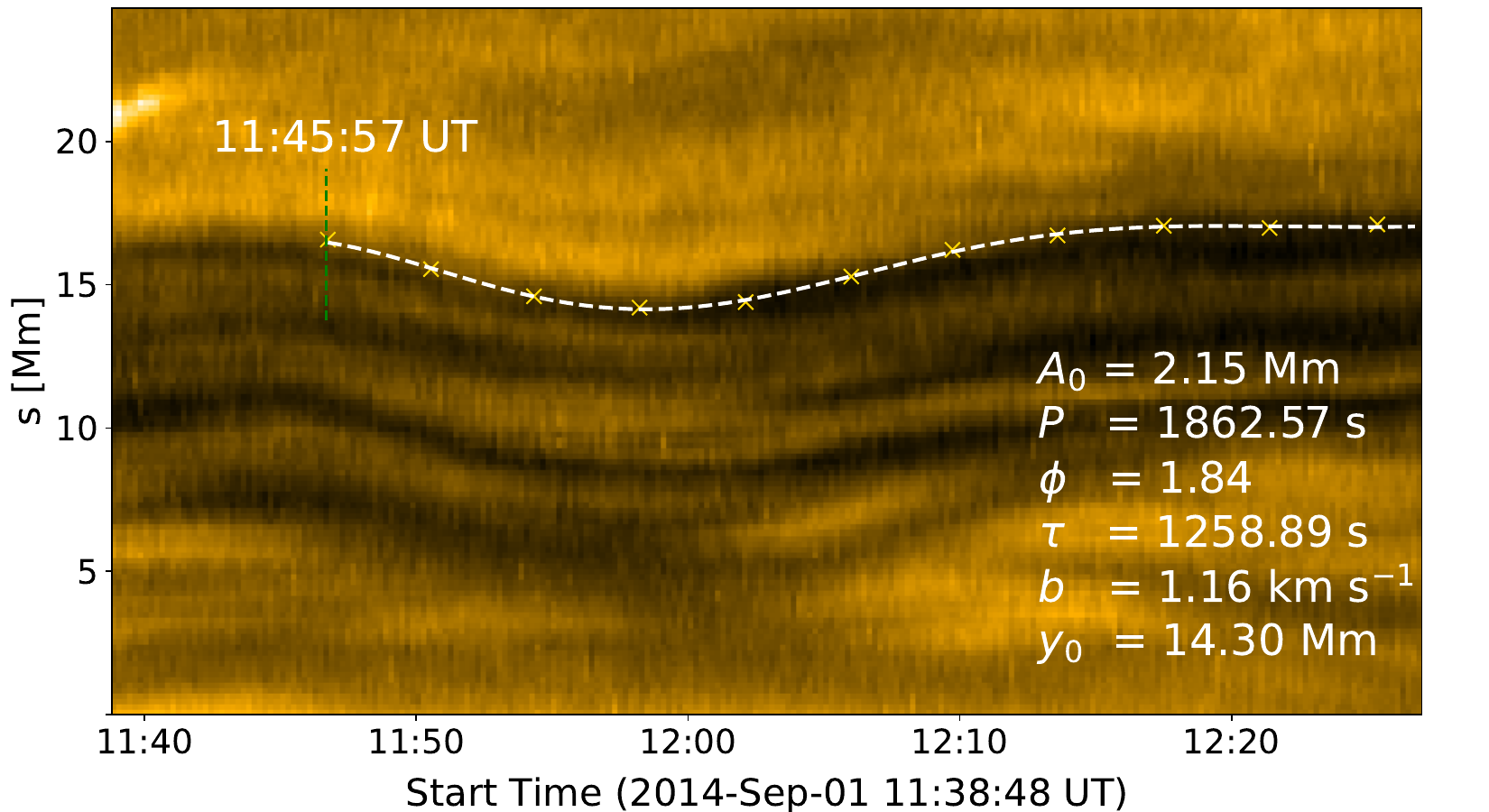}
\caption{{Time-distance diagram of slit F, showing the oscillation of
    the filament. The onset time of the oscillation was 11:45:57 UT.
    The dashed lines are fittings of the oscillation by using
    Equation~\ref{eqn-1}.
    An online animation of the AIA 171 {\AA} data illustrates the oscillation process of the filament. 
    This data is utilized to generate the time-distance diagram using slit F, as depicted in Figure~\ref{fig1}(c).
\label{fig11}}}
\end{figure}

To investigate the kinematics of the prominence, we selected three artificial slices in Figure~\ref{fig1}(c): S0, S1 and S2, which were equidistantly aligned with the direction of {vertical oscillatory component}.
All the slices are 10$\arcsec$ in width.
Time-distance diagrams are displayed in Figure~\ref{fig10}.
{The oscillation occurred at 11:15:33 UT and gradually ceased to oscillate around 13:00:00 UT, lasting for about three cycles.}
The four green points in panel (b2) represent the displacement of the prominence centroid obtained from the EUVI 304 {\AA} images using the radial-filter technique as detailed in Section~\ref{sec:dat}.
{These instances also correspond to the four temporal points when 3D reconstruction was conducted in Figure~\ref{fig9}.}.

The standard program \texttt{mpfit.pro} in the SSW is applied to determine the parameter of the oscillation by using the following fitting function:

\begin{equation}\label{eqn-1}
A(t)=A_0\sin(\frac{2\pi t}{P}+\Phi)e^{-\frac{t}{\tau}}+bt+y_0.
\end{equation}

where $A_0$, $P$, $\Phi$ and $\tau$ represent the initial amplitude, period, initial phase and damping time scale, respectively.
$bt+y_0$ represent a linear term of the equilibrium position of the prominence.
These parameters obtained are listed in Table~\ref{Tab1}.
The initial amplitudes ($A_0$) range from 14.76 to 18.40 Mm, with an average of 16.08 Mm. 
The initial heights ($y_0$) fall within the range of 38.56 to 62.05 Mm, with an average of 52.38 Mm.
The periods ($P$) and damping time scales ($\tau$) span from 1644 to 1882 seconds and 2545 to 2907 seconds, with mean values of 1769 seconds and 2640 seconds, respectively.
The calculated damping ratios ($\tau$/$P$) range from 1.35 to 1.68, with an average of 1.49.
The initial phases varies between 4.02 and 5.04 radians, with a mean value of 4.45 radians, suggesting oscillation in phase as a rigid body.
It should be noticed that the oscillation process is not well fitted by the function until after about 11:30 UT, when the prominence first reaches its height minimum.
One possible reason is that the prominence is not in equilibrium prior to the contact with the EUV wave.
Upon the onset of the downward oscillation of the prominence, it manifested a motion with a approximately uniform velocity.
Consequently, we performed a linear fitting on this process (as indicated by the white straight segment in Figure~\ref{fig10}), yielding velocities ranging from approximately -24.91 to -28.93 km s$^{-1}$, with an average velocity of around 27 km s$^{-1}$, directed towards the solar surface.

\begin{table*}
\setlength{\tabcolsep}{17pt}
\begin{center}
\caption{Fitted Parameters of the Prominence Oscillation.}
\label{Tab1}
 \begin{tabular}{ccccccccc}
  \hline\noalign{\smallskip}
passband & $A_0$  &  $P$     &  $\Phi$ &  $\tau$  &     $b$       & $y_0$  & $\tau$/$P$    \\
 ({\AA}) & (Mm)   &  (s)     &  (rad)  &   (s)    & (km s$^{-1}$) & (Mm)   &               \\
  \hline\noalign{\smallskip}
171 (S0)  & 16.78 & 1881.80  &  5.04   &  2544.67 &    0.93       &  62.05 &   1.35        \\
171 (S1)  & 14.92 & 1704.12  &  4.27   &  2587.68 &    -0.02      &  54.25 &   1.52        \\
171 (S2)  & 14.76 & 1730.92  &  4.36   &  2907.49 &    -2.60      &  43.73 &   1.68        \\
304 (S0)  & 15.53 & 1846.10  &  4.02   &  2645.27 &    1.30       &  61.32 &   1.43        \\
304 (S1)  & 16.08 & 1807.26  &  4.54   &  2584.54 &    -0.45      &  54.39 &   1.43        \\
304 (S2)  & 18.40 & 1643.82  &  4.46   &  2564.08 &    -0.02      &  38.56 &   1.56        \\
\hline
Average   & 16.08 & 1769.00  &  4.45   &  2639.96 &    -0.14      &  52.38 &   1.49        \\
  \noalign{\smallskip}\hline
\end{tabular}
\end{center}
\end{table*}

\begin{table*}
\setlength{\tabcolsep}{17pt}
\begin{center}
\caption{Fitted Parameters of the filament Oscillation.}
\label{Tab2}
 \begin{tabular}{ccccccccc}
  \hline\noalign{\smallskip}
passband & $A_0$  &  $P$     &  $\Phi$ &  $\tau$  &     $b$       & $y_0$  & $\tau$/$P$    \\
 ({\AA}) & (Mm)   &  (s)     &  (rad)  &   (s)    & (km s$^{-1}$) & (Mm)   &               \\
  \hline\noalign{\smallskip}
171 (F)  & 2.15 & 1862.57  &  1.84   &  1258.59 &    1.16       &  14.30 &    0.68       \\
  \noalign{\smallskip}\hline
\end{tabular}
\end{center}
\end{table*}

{The filament exhibits faint transverse oscillations in the AIA field of view, approximately over one cycle, as depicted in the online animation associated with Figure~\ref{fig11}.
Due to the exclusive observation by AIA, 3D reconstruction is not feasible.
Therefore, the selected slice F for analysis are shown in Figure~\ref{fig1}(c), and the resulting time-distance diagram is shown in Figure~\ref{fig11} (We chose AIA 171 {\AA} data because where oscillatory motion is more pronounced).
Similarly, we employed equation~\ref{eqn-1} to fit the result, and the obtained parameters are presented in the Table~\ref{Tab2}.
}

\section{Discussion} \label{sec:Dis}

\subsection{EUV wave}
The EUV wave investigated in this study was generated by {CME} occurring on the farside of the Sun and thus was observed by STB.
The EUV wave propagated in all directions but was obstructed when encountering magnetic structures.
Specifically, to the southeast of the flare site, there existed an elongated magnetic structure.
As the EUV wave interacted with this structure, it generated a standing wave.
In addition, a subtle wave passing through the magnetic structure can be observed in the base-difference movie, affirming the fundamental wave nature.

Owing to the limited time cadence of the EUVI in this period (5 minutes for 195 {\AA} and 10 minutes for 304 {\AA}), the properties of the EUV waves can not be investigated in detail.
{Alternatively, by determining the positions of the flare, prominence, and filament, we calculated the linear speed of the EUV wave.
The speed of the EUV wave reaching the prominence was 498 km s$^{-1}$, while reaching the filament, it was 451 km s$^{-1}$.
This is in accordance with the speed characteristics of EUV waves \citep{Asai2012,Tak2015}.
Interestingly, even after covering such a long distance (605 Mm to 1351 Mm), the speed of the EUV wave did not significantly decrease.
In the upward direction, the CME attained a maximum speed of 1901 km s$^{-1}$, with a average speed of 650 km s$^{-1}$, and the velocity of the driven shock wave also reached 814 km s$^{-1}$.
On the other hand, the coronal dimming got darker but with only little (or no) further expansion.
These observational phenomena indicate that the upward shock is driven by the CME, whereas in the lateral direction, the wave freely propagates once the lateral expansion of the CME flanks stops \citep{Ver2010,She2012b}.
The occurrence of a type II radio burst further confirms the fast-mode nature of the EUV wave.
}

\subsection{The oscillation of the prominence and filament}
{The prominence oscillation was simultaneously observed by both STB and SDO, which is unlikely to have commenced oscillation from an equilibrium state, as its initial trajectory did not align with subsequent motion that can be adequately fit by a single sinusoidal function.
This dual perspective enables us to perform a 3D reconstruction to the filament.
As shown in Figure~\ref{fig9}, the prominence exhibits simultaneous horizontal and vertical oscillations, indicative of transverse oscillations. 
This implies that the normal vector of the wave is inclined downward toward the prominence \citep{Tak2015,Zhe2023,Dai2023}, which appears to provide strong support for discussion on the properties of EUV wave above (can also see the Figure 9 in \cite{She2014b}).}

The main restoring force of prominence oscillation is generally attributed to magnetic tension by using a slab model \citep{Dia2001,Zha2018,Zho2018}.
Considering that the periods of oscillation at different positions are approximately equal (see Table~\ref{Tab1}), we believe that the prominence experienced a global {transverse} oscillation of fast kink mode \citep{Zho2016}.
The magnetic field strength of the prominence can be roughly estimated as follows \citep{Hyd1966}:

\begin{equation} \label{eqn-2}
  B_r^2 = \pi \rho_p r_0^2(4\pi^2P^{-2}+\tau^{-2}).
\end{equation}

where $B_r$, $\rho_p$ and $r_0$ represent the radial component of magnetic field, density, and scale height of the prominence, respectively.
P and $\tau$ are the period and damping time scale of the oscillation.
r$_0$ is approximately $4.3 \times 10^9 $ cm based on {the 3D reconstruction}.
By taking the average values of the measured period and decay time as 1769 s and 2640 s (see Table~\ref{Tab1}), respectively, and assuming $\rho_p = 4 \times 10^{-14}$ g cm$^{-3}$, we can estimate $B_r$ to be 5.4 G.
{Likewise, the radial magnetic field of the filament can also be calculated, yielding an estimate of 4.1 G.}
It is close to the values reported in previous papers \citep{Hyd1966,She2014,She2017,Zha2018,Dai2023}.

\section{Summary} \label{sec:Sum}
In this paper, we report our multi-angle observations of the transverse oscillation of a prominence {and a filament sequentially}, which were induced by an EUV wave on 2014 September 1.
The main results are summarized as follows:

\begin{enumerate}
\item{{The upward shock was propelled by the CME, whereas in the lateral direction, the EUV wave freely propagated upon the expansion of the CME flanks stopped, which can be supported by the coronal dimming region with little (or no) expansion.
The EUV wave propagated over such an extended distance (605 Mm to 1351 Mm) with minimal speed reduction (498 km s$^{-1}$ to 451 km s$^{-1}$).
Moreover, the occurrence of type II radio bursts further confirmed the EUV wave as a fast-mode shock.}
}
\item{{The EUV wave induced oscillations in both a prominence and a filament, sequentially.
3D reconstruction reveals that the prominence exhibits both vertical and horizontal oscillations, suggesting a downward inclination of the normal vector of the EUV wave. 
The angle to the local photosphere plane can be further calculated to be approximately 63$^\circ$.
This is consistent with the velocity characteristics of the EUV wave discussed above. 
The oscillations of the prominence and filament also allow for the determination of their radial magnetic field strengths, approximated at 5.4 G and 4.1 G, respectively.}
}
\end{enumerate}

\begin{acknowledgments}
STEREO/SECCHI data are provided by a consortium of US, UK, Germany, Belgium, and France.
SDO is a mission of NASA\rq{}s Living With a Star Program. AIA and HMI data are courtesy of the NASA/SDO science teams.
The e-Callisto data are courtesy of the Institute for Data Science FHNW Brugg/Windisch, Switzerland.
This work is supported by the National Key R$\&$D Program of China 2021YFA1600500 (2021YFA1600502), NSFC under grant 12373065, 
and Yunnan Key Laboratory of Solar Physics and Space Science under the number YNSPCC202206.
\end{acknowledgments}

\bibliography{sample631}{}

\begin{thebibliography}{}
\expandafter\ifx\csname natexlab\endcsname\relax\def\natexlab#1{#1}\fi
\providecommand{\url}[1]{\href{#1}{#1}}
\providecommand{\dodoi}[1]{doi:~\href{http://doi.org/#1}{\nolinkurl{#1}}}
\providecommand{\doeprint}[1]{\href{http://ascl.net/#1}{\nolinkurl{http://ascl.net/#1}}}
\providecommand{\doarXiv}[1]{\href{https://arxiv.org/abs/#1}{\nolinkurl{https://arxiv.org/abs/#1}}}

\bibitem[{{Asai} {et~al.}(2012){Asai}, {Ishii}, {Isobe}, {Kitai}, {Ichimoto},
  {UeNo}, {Nagata}, {Morita}, {Nishida}, {Shiota}, {Oi}, {Akioka}, \&
  {Shibata}}]{Asai2012}
{Asai}, A., {Ishii}, T.~T., {Isobe}, H., {et~al.} 2012, \apjl, 745, L18,
  \dodoi{10.1088/2041-8205/745/2/L18}

\bibitem[{{Brueckner} {et~al.}(1995){Brueckner}, {Howard}, {Koomen},
  {Korendyke}, {Michels}, {Moses}, {Socker}, {Dere}, {Lamy}, {Llebaria},
  {Bout}, {Schwenn}, {Simnett}, {Bedford}, \& {Eyles}}]{Bru1995}
{Brueckner}, G.~E., {Howard}, R.~A., {Koomen}, M.~J., {et~al.} 1995, \solphys,
  162, 357, \dodoi{10.1007/BF00733434}

\bibitem[{{Chen}(2016)}]{Che2016}
{Chen}, P.~F. 2016, Washington DC American Geophysical Union Geophysical
  Monograph Series, 216, 381, \dodoi{10.1002/9781119055006.ch22}

\bibitem[{{Chen} \& {Wu}(2011)}]{Che2011}
{Chen}, P.~F., \& {Wu}, Y. 2011, \apjl, 732, L20,
  \dodoi{10.1088/2041-8205/732/2/L20}

\bibitem[{{Chen} {et~al.}(2020){Chen}, {Xu}, \& {Ding}}]{Che2020}
{Chen}, P.-F., {Xu}, A.-A., \& {Ding}, M.-D. 2020, Research in Astronomy and
  Astrophysics, 20, 166, \dodoi{10.1088/1674-4527/20/10/166}

\bibitem[{{Dai} {et~al.}(2023){Dai}, {Zhang}, {Qiu}, {Li}, {Li}, {Li}, {Su}, \&
  {Ji}}]{Dai2023}
{Dai}, J., {Zhang}, Q., {Qiu}, Y., {et~al.} 2023, \apj, 959, 71,
  \dodoi{10.3847/1538-4357/ad0839}

\bibitem[{{Dai} {et~al.}(2021){Dai}, {Zhang}, {Zhang}, {Xu}, {Su}, \&
  {Ji}}]{Dai2021}
{Dai}, J., {Zhang}, Q., {Zhang}, Y., {et~al.} 2021, \apj, 923, 74,
  \dodoi{10.3847/1538-4357/ac2d97}

\bibitem[{{Delaboudini{\`e}re} {et~al.}(1995){Delaboudini{\`e}re}, {Artzner},
  {Brunaud}, {Gabriel}, {Hochedez}, {Millier}, {Song}, {Au}, {Dere}, {Howard},
  {Kreplin}, {Michels}, {Moses}, {Defise}, {Jamar}, {Rochus}, {Chauvineau},
  {Marioge}, {Catura}, {Lemen}, {Shing}, {Stern}, {Gurman}, {Neupert},
  {Maucherat}, {Clette}, {Cugnon}, \& {Van Dessel}}]{Del1995}
{Delaboudini{\`e}re}, J.~P., {Artzner}, G.~E., {Brunaud}, J., {et~al.} 1995,
  \solphys, 162, 291, \dodoi{10.1007/BF00733432}

\bibitem[{{Devi} {et~al.}(2022){Devi}, {Chandra}, {Joshi}, {Chen}, {Schmieder},
  {Uddin}, \& {Moon}}]{Dev2022}
{Devi}, P., {Chandra}, R., {Joshi}, R., {et~al.} 2022, Advances in Space
  Research, 70, 1592, \dodoi{10.1016/j.asr.2022.02.053}

\bibitem[{{D{\'\i}az} {et~al.}(2001){D{\'\i}az}, {Oliver}, {Erd{\'e}lyi}, \&
  {Ballester}}]{Dia2001}
{D{\'\i}az}, A.~J., {Oliver}, R., {Erd{\'e}lyi}, R., \& {Ballester}, J.~L.
  2001, \aap, 379, 1083, \dodoi{10.1051/0004-6361:20011351}

\bibitem[{{Eto} {et~al.}(2002){Eto}, {Isobe}, {Narukage}, {Asai}, {Morimoto},
  {Thompson}, {Yashiro}, {Wang}, {Kitai}, {Kurokawa}, \& {Shibata}}]{Eto2002}
{Eto}, S., {Isobe}, H., {Narukage}, N., {et~al.} 2002, \pasj, 54, 481,
  \dodoi{10.1093/pasj/54.3.481}

\bibitem[{{Gallagher} \& {Long}(2011)}]{Gal2011}
{Gallagher}, P.~T., \& {Long}, D.~M. 2011, \ssr, 158, 365,
  \dodoi{10.1007/s11214-010-9710-7}

\bibitem[{{Heinzel} {et~al.}(2008){Heinzel}, {Schmieder}, {F{\'a}rn{\'\i}k},
  {Schwartz}, {Labrosse}, {Kotr{\v{c}}}, {Anzer}, {Molodij}, {Berlicki},
  {DeLuca}, {Golub}, {Watanabe}, \& {Berger}}]{Hei2008}
{Heinzel}, P., {Schmieder}, B., {F{\'a}rn{\'\i}k}, F., {et~al.} 2008, \apj,
  686, 1383, \dodoi{10.1086/591018}

\bibitem[{{Howard} {et~al.}(2008){Howard}, {Moses}, {Vourlidas}, {Newmark},
  {Socker}, {Plunkett}, {Korendyke}, {Cook}, {Hurley}, {Davila}, {Thompson},
  {St Cyr}, {Mentzell}, {Mehalick}, {Lemen}, {Wuelser}, {Duncan}, {Tarbell},
  {Wolfson}, {Moore}, {Harrison}, {Waltham}, {Lang}, {Davis}, {Eyles},
  {Mapson-Menard}, {Simnett}, {Halain}, {Defise}, {Mazy}, {Rochus}, {Mercier},
  {Ravet}, {Delmotte}, {Auchere}, {Delaboudiniere}, {Bothmer}, {Deutsch},
  {Wang}, {Rich}, {Cooper}, {Stephens}, {Maahs}, {Baugh}, {McMullin}, \&
  {Carter}}]{How2008}
{Howard}, R.~A., {Moses}, J.~D., {Vourlidas}, A., {et~al.} 2008, \ssr, 136, 67,
  \dodoi{10.1007/s11214-008-9341-4}

\bibitem[{{Hyder}(1966)}]{Hyd1966}
{Hyder}, C.~L. 1966, \zap, 63, 78

\bibitem[{{Jing} {et~al.}(2003){Jing}, {Lee}, {Spirock}, {Xu}, {Wang}, \&
  {Choe}}]{Jin2003}
{Jing}, J., {Lee}, J., {Spirock}, T.~J., {et~al.} 2003, \apjl, 584, L103,
  \dodoi{10.1086/373886}

\bibitem[{{Kaiser} {et~al.}(2008){Kaiser}, {Kucera}, {Davila}, {St. Cyr},
  {Guhathakurta}, \& {Christian}}]{Kai2008}
{Kaiser}, M.~L., {Kucera}, T.~A., {Davila}, J.~M., {et~al.} 2008, \ssr, 136, 5,
  \dodoi{10.1007/s11214-007-9277-0}

\bibitem[{{Kleczek} \& {Kuperus}(1969)}]{Kle1969}
{Kleczek}, J., \& {Kuperus}, M. 1969, \solphys, 6, 72,
  \dodoi{10.1007/BF00146797}

\bibitem[{{Krucker} {et~al.}(2011){Krucker}, {Kontar}, {Christe}, {Glesener},
  \& {Lin}}]{Kru2011}
{Krucker}, S., {Kontar}, E.~P., {Christe}, S., {Glesener}, L., \& {Lin}, R.~P.
  2011, \apj, 742, 82, \dodoi{10.1088/0004-637X/742/2/82}

\bibitem[{{Labrosse} {et~al.}(2010){Labrosse}, {Heinzel}, {Vial}, {Kucera},
  {Parenti}, {Gun{\'a}r}, {Schmieder}, \& {Kilper}}]{Lab2010}
{Labrosse}, N., {Heinzel}, P., {Vial}, J.~C., {et~al.} 2010, \ssr, 151, 243,
  \dodoi{10.1007/s11214-010-9630-6}

\bibitem[{{Lemen} {et~al.}(2012){Lemen}, {Title}, {Akin}, {Boerner}, {Chou},
  {Drake}, {Duncan}, {Edwards}, {Friedlaender}, {Heyman}, {Hurlburt}, {Katz},
  {Kushner}, {Levay}, {Lindgren}, {Mathur}, {McFeaters}, {Mitchell}, {Rehse},
  {Schrijver}, {Springer}, {Stern}, {Tarbell}, {Wuelser}, {Wolfson}, {Yanari},
  {Bookbinder}, {Cheimets}, {Caldwell}, {Deluca}, {Gates}, {Golub}, {Park},
  {Podgorski}, {Bush}, {Scherrer}, {Gummin}, {Smith}, {Auker}, {Jerram},
  {Pool}, {Soufli}, {Windt}, {Beardsley}, {Clapp}, {Lang}, \&
  {Waltham}}]{Lem2012}
{Lemen}, J.~R., {Title}, A.~M., {Akin}, D.~J., {et~al.} 2012, \solphys, 275,
  17, \dodoi{10.1007/s11207-011-9776-8}

\bibitem[{{Li} \& {Zhang}(2012)}]{Li2012}
{Li}, T., \& {Zhang}, J. 2012, \apjl, 760, L10,
  \dodoi{10.1088/2041-8205/760/1/L10}

\bibitem[{{Liakh} {et~al.}(2023){Liakh}, {Luna}, \& {Khomenko}}]{Lia2023}
{Liakh}, V., {Luna}, M., \& {Khomenko}, E. 2023, \aap, 673, A154,
  \dodoi{10.1051/0004-6361/202245765}

\bibitem[{{Liu} \& {Ofman}(2014)}]{Liu2014}
{Liu}, W., \& {Ofman}, L. 2014, \solphys, 289, 3233,
  \dodoi{10.1007/s11207-014-0528-4}

\bibitem[{{Luna} \& {Karpen}(2012)}]{Lun2012}
{Luna}, M., \& {Karpen}, J. 2012, \apjl, 750, L1,
  \dodoi{10.1088/2041-8205/750/1/L1}

\bibitem[{{Luna} {et~al.}(2014){Luna}, {Knizhnik}, {Muglach}, {Karpen},
  {Gilbert}, {Kucera}, \& {Uritsky}}]{Lun2014}
{Luna}, M., {Knizhnik}, K., {Muglach}, K., {et~al.} 2014, \apj, 785, 79,
  \dodoi{10.1088/0004-637X/785/1/79}

\bibitem[{{Luna} {et~al.}(2016){Luna}, {Terradas}, {Khomenko}, {Collados}, \&
  {de Vicente}}]{Lun2016}
{Luna}, M., {Terradas}, J., {Khomenko}, E., {Collados}, M., \& {de Vicente}, A.
  2016, \apj, 817, 157, \dodoi{10.3847/0004-637X/817/2/157}

\bibitem[{{Mancuso} {et~al.}(2019){Mancuso}, {Frassati}, {Bemporad}, \&
  {Barghini}}]{Man2019}
{Mancuso}, S., {Frassati}, F., {Bemporad}, A., \& {Barghini}, D. 2019, \aap,
  624, L2, \dodoi{10.1051/0004-6361/201935157}

\bibitem[{{Martin}(1998)}]{Mar1998}
{Martin}, S.~F. 1998, \solphys, 182, 107, \dodoi{10.1023/A:1005026814076}

\bibitem[{{Masson} {et~al.}(2013){Masson}, {Antiochos}, \& {DeVore}}]{Mas2013}
{Masson}, S., {Antiochos}, S.~K., \& {DeVore}, C.~R. 2013, \apj, 771, 82,
  \dodoi{10.1088/0004-637X/771/2/82}

\bibitem[{{Moreton}(1960)}]{Mor1960}
{Moreton}, G.~E. 1960, \aj, 65, 494, \dodoi{10.1086/108346}

\bibitem[{{Muhr} {et~al.}(2010){Muhr}, {Vr{\v{s}}nak}, {Temmer}, {Veronig}, \&
  {Magdaleni{\'c}}}]{Muh2010}
{Muhr}, N., {Vr{\v{s}}nak}, B., {Temmer}, M., {Veronig}, A.~M., \&
  {Magdaleni{\'c}}, J. 2010, \apj, 708, 1639,
  \dodoi{10.1088/0004-637X/708/2/1639}

\bibitem[{{Ofman} \& {Thompson}(2002)}]{Ofm2002}
{Ofman}, L., \& {Thompson}, B.~J. 2002, \apj, 574, 440, \dodoi{10.1086/340924}

\bibitem[{{Okamoto} {et~al.}(2004){Okamoto}, {Nakai}, {Keiyama}, {Narukage},
  {UeNo}, {Kitai}, {Kurokawa}, \& {Shibata}}]{Oka2004}
{Okamoto}, T.~J., {Nakai}, H., {Keiyama}, A., {et~al.} 2004, \apj, 608, 1124,
  \dodoi{10.1086/420838}

\bibitem[{{Ontiveros} \& {Vourlidas}(2009)}]{On2009}
{Ontiveros}, V., \& {Vourlidas}, A. 2009, \apj, 693, 267,
  \dodoi{10.1088/0004-637X/693/1/267}

\bibitem[{{Patsourakos} \& {Vourlidas}(2009)}]{Pat2009}
{Patsourakos}, S., \& {Vourlidas}, A. 2009, \apjl, 700, L182,
  \dodoi{10.1088/0004-637X/700/2/L182}

\bibitem[{{Patsourakos} \& {Vourlidas}(2012)}]{Pat2012}
---. 2012, \solphys, 281, 187, \dodoi{10.1007/s11207-012-9988-6}

\bibitem[{{Pesnell} {et~al.}(2012){Pesnell}, {Thompson}, \&
  {Chamberlin}}]{Pes2012}
{Pesnell}, W.~D., {Thompson}, B.~J., \& {Chamberlin}, P.~C. 2012, \solphys,
  275, 3, \dodoi{10.1007/s11207-011-9841-3}

\bibitem[{{Ramsey} \& {Smith}(1966)}]{Ram1966}
{Ramsey}, H.~E., \& {Smith}, S.~F. 1966, \aj, 71, 197, \dodoi{10.1086/109903}

\bibitem[{{Shen} {et~al.}(2014{\natexlab{a}}){Shen}, {Ichimoto}, {Ishii},
  {Tian}, {Zhao}, \& {Shibata}}]{She2014}
{Shen}, Y., {Ichimoto}, K., {Ishii}, T.~T., {et~al.} 2014{\natexlab{a}}, \apj,
  786, 151, \dodoi{10.1088/0004-637X/786/2/151}

\bibitem[{{Shen} \& {Liu}(2012{\natexlab{a}})}]{She2012b}
{Shen}, Y., \& {Liu}, Y. 2012{\natexlab{a}}, \apjl, 752, L23,
  \dodoi{10.1088/2041-8205/752/2/L23}

\bibitem[{{Shen} \& {Liu}(2012{\natexlab{b}})}]{She2012}
---. 2012{\natexlab{b}}, \apj, 754, 7, \dodoi{10.1088/0004-637X/754/1/7}

\bibitem[{{Shen} {et~al.}(2017){Shen}, {Liu}, {Tian}, \& {Qu}}]{She2017}
{Shen}, Y., {Liu}, Y., {Tian}, Z., \& {Qu}, Z. 2017, \apj, 851, 101,
  \dodoi{10.3847/1538-4357/aa9af0}

\bibitem[{{Shen} {et~al.}(2014{\natexlab{b}}){Shen}, {Liu}, {Chen}, \&
  {Ichimoto}}]{She2014b}
{Shen}, Y., {Liu}, Y.~D., {Chen}, P.~F., \& {Ichimoto}, K. 2014{\natexlab{b}},
  \apj, 795, 130, \dodoi{10.1088/0004-637X/795/2/130}

\bibitem[{{Shen} {et~al.}(2019){Shen}, {Qu}, {Yuan}, {Chen}, {Duan}, {Zhou},
  {Tang}, {Huang}, \& {Liu}}]{She2019}
{Shen}, Y., {Qu}, Z., {Yuan}, D., {et~al.} 2019, \apj, 883, 104,
  \dodoi{10.3847/1538-4357/ab3a4d}

\bibitem[{{Takahashi} {et~al.}(2015){Takahashi}, {Asai}, \&
  {Shibata}}]{Tak2015}
{Takahashi}, T., {Asai}, A., \& {Shibata}, K. 2015, \apj, 801, 37,
  \dodoi{10.1088/0004-637X/801/1/37}

\bibitem[{{Tan} {et~al.}(2023){Tan}, {Shen}, {Zhou}, {Tang}, {Zhou}, {Duan}, \&
  {Yao}}]{Tan2023}
{Tan}, S., {Shen}, Y., {Zhou}, X., {et~al.} 2023, \mnras, 520, 3080,
  \dodoi{10.1093/mnras/stad295}

\bibitem[{{Tandberg-Hanssen}(1995)}]{Tan1995}
{Tandberg-Hanssen}, E. 1995, Science, 269, 111

\bibitem[{{Temmer} {et~al.}(2011){Temmer}, {Veronig}, {Gopalswamy}, \&
  {Yashiro}}]{Tem2011}
{Temmer}, M., {Veronig}, A.~M., {Gopalswamy}, N., \& {Yashiro}, S. 2011,
  \solphys, 273, 421, \dodoi{10.1007/s11207-011-9746-1}

\bibitem[{{Terradas} {et~al.}(2013){Terradas}, {Soler}, {D{\'\i}az}, {Oliver},
  \& {Ballester}}]{Ter2013}
{Terradas}, J., {Soler}, R., {D{\'\i}az}, A.~J., {Oliver}, R., \& {Ballester},
  J.~L. 2013, \apj, 778, 49, \dodoi{10.1088/0004-637X/778/1/49}

\bibitem[{{Terradas} {et~al.}(2015){Terradas}, {Soler}, {Luna}, {Oliver}, \&
  {Ballester}}]{Ter2015}
{Terradas}, J., {Soler}, R., {Luna}, M., {Oliver}, R., \& {Ballester}, J.~L.
  2015, \apj, 799, 94, \dodoi{10.1088/0004-637X/799/1/94}

\bibitem[{{Thompson} {et~al.}(1998){Thompson}, {Plunkett}, {Gurman}, {Newmark},
  {St. Cyr}, \& {Michels}}]{Tho1998}
{Thompson}, B.~J., {Plunkett}, S.~P., {Gurman}, J.~B., {et~al.} 1998, \grl, 25,
  2465, \dodoi{10.1029/98GL50429}

\bibitem[{{Thompson} {et~al.}(2000){Thompson}, {Reynolds}, {Aurass},
  {Gopalswamy}, {Gurman}, {Hudson}, {Martin}, \& {St. Cyr}}]{Tho2000}
{Thompson}, B.~J., {Reynolds}, B., {Aurass}, H., {et~al.} 2000, \solphys, 193,
  161, \dodoi{10.1023/A:1005222123970}

\bibitem[{{Thompson} {et~al.}(1999){Thompson}, {Gurman}, {Neupert}, {Newmark},
  {Delaboudini{\`e}re}, {Cyr}, {Stezelberger}, {Dere}, {Howard}, \&
  {Michels}}]{Tho1999}
{Thompson}, B.~J., {Gurman}, J.~B., {Neupert}, W.~M., {et~al.} 1999, \apjl,
  517, L151, \dodoi{10.1086/312030}

\bibitem[{{Tripathi} \& {Raouafi}(2007)}]{Tri2007}
{Tripathi}, D., \& {Raouafi}, N.~E. 2007, \aap, 473, 951,
  \dodoi{10.1051/0004-6361:20077255}

\bibitem[{{Uchida}(1968)}]{Uch1968}
{Uchida}, Y. 1968, \solphys, 4, 30, \dodoi{10.1007/BF00146996}

\bibitem[{{Uchida}(1974)}]{Uch1974}
---. 1974, \solphys, 39, 431, \dodoi{10.1007/BF00162436}

\bibitem[{{Veronig} {et~al.}(2010){Veronig}, {Muhr}, {Kienreich}, {Temmer}, \&
  {Vr{\v{s}}nak}}]{Ver2010}
{Veronig}, A.~M., {Muhr}, N., {Kienreich}, I.~W., {Temmer}, M., \&
  {Vr{\v{s}}nak}, B. 2010, \apjl, 716, L57, \dodoi{10.1088/2041-8205/716/1/L57}

\bibitem[{{Vr{\v{s}}nak} {et~al.}(2002){Vr{\v{s}}nak}, {Warmuth},
  {Braj{\v{s}}a}, \& {Hanslmeier}}]{Vrs2002}
{Vr{\v{s}}nak}, B., {Warmuth}, A., {Braj{\v{s}}a}, R., \& {Hanslmeier}, A.
  2002, \aap, 394, 299, \dodoi{10.1051/0004-6361:20021121}

\bibitem[{{Wang}(2000)}]{Wan2000}
{Wang}, Y.~M. 2000, \apjl, 543, L89, \dodoi{10.1086/318178}

\bibitem[{{Warmuth}(2007)}]{War2007}
{Warmuth}, A. 2007, in Lecture Notes in Physics, Berlin Springer Verlag, ed.
  K.-L. {Klein} \& A.~L. {MacKinnon}, Vol. 725, 107

\bibitem[{{Warmuth}(2015)}]{War2015}
---. 2015, Living Reviews in Solar Physics, 12, 3, \dodoi{10.1007/lrsp-2015-3}

\bibitem[{{Warmuth} {et~al.}(2001){Warmuth}, {Vr{\v{s}}nak}, {Aurass}, \&
  {Hanslmeier}}]{War2001}
{Warmuth}, A., {Vr{\v{s}}nak}, B., {Aurass}, H., \& {Hanslmeier}, A. 2001,
  \apjl, 560, L105, \dodoi{10.1086/324055}

\bibitem[{{Wills-Davey} \& {Attrill}(2009)}]{Wil2009}
{Wills-Davey}, M.~J., \& {Attrill}, G.~D.~R. 2009, \ssr, 149, 325,
  \dodoi{10.1007/s11214-009-9612-8}

\bibitem[{{Wu} {et~al.}(2001){Wu}, {Zheng}, {Wang}, {Thompson}, {Plunkett},
  {Zhao}, \& {Dryer}}]{Wu2001}
{Wu}, S.~T., {Zheng}, H., {Wang}, S., {et~al.} 2001, \jgr, 106, 25089,
  \dodoi{10.1029/2000JA000447}

\bibitem[{{Wuelser} {et~al.}(2004){Wuelser}, {Lemen}, {Tarbell}, {Wolfson},
  {Cannon}, {Carpenter}, {Duncan}, {Gradwohl}, {Meyer}, {Moore}, {Navarro},
  {Pearson}, {Rossi}, {Springer}, {Howard}, {Moses}, {Newmark},
  {Delaboudiniere}, {Artzner}, {Auchere}, {Bougnet}, {Bouyries}, {Bridou},
  {Clotaire}, {Colas}, {Delmotte}, {Jerome}, {Lamare}, {Mercier}, {Mullot},
  {Ravet}, {Song}, {Bothmer}, \& {Deutsch}}]{Wue2004}
{Wuelser}, J.-P., {Lemen}, J.~R., {Tarbell}, T.~D., {et~al.} 2004, in Society
  of Photo-Optical Instrumentation Engineers (SPIE) Conference Series, Vol.
  5171, Telescopes and Instrumentation for Solar Astrophysics, ed.
  S.~{Fineschi} \& M.~A. {Gummin}, 111--122, \dodoi{10.1117/12.506877}

\bibitem[{{Wyper} {et~al.}(2018){Wyper}, {DeVore}, \& {Antiochos}}]{Wyp2018}
{Wyper}, P.~F., {DeVore}, C.~R., \& {Antiochos}, S.~K. 2018, \apj, 852, 98,
  \dodoi{10.3847/1538-4357/aa9ffc}

\bibitem[{{Yan} {et~al.}(2015){Yan}, {Xue}, {Pan}, {Wang}, {Xiang}, {Kong}, \&
  {Yang}}]{Yan2015}
{Yan}, X.~L., {Xue}, Z.~K., {Pan}, G.~M., {et~al.} 2015, \apjs, 219, 17,
  \dodoi{10.1088/0067-0049/219/2/17}

\bibitem[{{Yashiro} {et~al.}(2008){Yashiro}, {Michalek}, \&
  {Gopalswamy}}]{Yas2008}
{Yashiro}, S., {Michalek}, G., \& {Gopalswamy}, N. 2008, Annales Geophysicae,
  26, 3103, \dodoi{10.5194/angeo-26-3103-2008}

\bibitem[{{Zhang} {et~al.}(2012){Zhang}, {Chen}, {Xia}, \& {Keppens}}]{Zha2012}
{Zhang}, Q.~M., {Chen}, P.~F., {Xia}, C., \& {Keppens}, R. 2012, \aap, 542,
  A52, \dodoi{10.1051/0004-6361/201218786}

\bibitem[{{Zhang} {et~al.}(2013){Zhang}, {Chen}, {Xia}, {Keppens}, \&
  {Ji}}]{Zha2013}
{Zhang}, Q.~M., {Chen}, P.~F., {Xia}, C., {Keppens}, R., \& {Ji}, H.~S. 2013,
  \aap, 554, A124, \dodoi{10.1051/0004-6361/201220705}

\bibitem[{{Zhang} \& {Ji}(2018)}]{Zha2018}
{Zhang}, Q.~M., \& {Ji}, H.~S. 2018, \apj, 860, 113,
  \dodoi{10.3847/1538-4357/aac37e}

\bibitem[{{Zhang} {et~al.}(2015){Zhang}, {Ning}, {Guo}, {Zhou}, {Cheng}, {Ji},
  {Feng}, \& {Wiegelmann}}]{Zha2015}
{Zhang}, Q.~M., {Ning}, Z.~J., {Guo}, Y., {et~al.} 2015, \apj, 805, 4,
  \dodoi{10.1088/0004-637X/805/1/4}

\bibitem[{{Zhang} {et~al.}(2011){Zhang}, {Kitai}, {Narukage}, {Matsumoto},
  {Ueno}, {Shibata}, \& {Wang}}]{Zha2011}
{Zhang}, Y., {Kitai}, R., {Narukage}, N., {et~al.} 2011, \pasj, 63, 685,
  \dodoi{10.1093/pasj/63.3.685}

\bibitem[{{Zheng} {et~al.}(2023){Zheng}, {Liu}, {Liu}, {Wang}, {Hou}, {Feng},
  {Kong}, {Huang}, {Song}, {Tian}, {Chen}, {Erd{\'e}lyi}, \& {Chen}}]{Zhe2023}
{Zheng}, R., {Liu}, Y., {Liu}, W., {et~al.} 2023, \apjl, 949, L8,
  \dodoi{10.3847/2041-8213/acd0ac}

\bibitem[{{Zhou} {et~al.}(2016){Zhou}, {Zhang}, \& {Wang}}]{Zho2016}
{Zhou}, G.~P., {Zhang}, J., \& {Wang}, J.~X. 2016, \apjl, 823, L19,
  \dodoi{10.3847/2041-8205/823/1/L19}

\bibitem[{{Zhou} {et~al.}(2018){Zhou}, {Xia}, {Keppens}, {Fang}, \&
  {Chen}}]{Zho2018}
{Zhou}, Y.-H., {Xia}, C., {Keppens}, R., {Fang}, C., \& {Chen}, P.~F. 2018,
  \apj, 856, 179, \dodoi{10.3847/1538-4357/aab614}

\bibitem[{{Zhou} {et~al.}(2017){Zhou}, {Zhang}, {Ouyang}, {Chen}, \&
  {Fang}}]{Zho2017}
{Zhou}, Y.-H., {Zhang}, L.-Y., {Ouyang}, Y., {Chen}, P.~F., \& {Fang}, C. 2017,
  \apj, 839, 9, \dodoi{10.3847/1538-4357/aa67de}

\bibitem[{{Zucca} {et~al.}(2018){Zucca}, {Morosan}, {Rouillard}, {Fallows},
  {Gallagher}, {Magdalenic}, {Klein}, {Mann}, {Vocks}, {Carley}, {Bisi},
  {Kontar}, {Rothkaehl}, {Dabrowski}, {Krankowski}, {Anderson}, {Asgekar},
  {Bell}, {Bentum}, {Best}, {Blaauw}, {Breitling}, {Broderick}, {Brouw},
  {Br{\"u}ggen}, {Butcher}, {Ciardi}, {de Geus}, {Deller}, {Duscha},
  {Eisl{\"o}ffel}, {Garrett}, {Grie{\ss}meier}, {Gunst}, {Heald}, {Hoeft},
  {H{\"o}randel}, {Iacobelli}, {Juette}, {Karastergiou}, {van Leeuwen},
  {McKay-Bukowski}, {Mulder}, {Munk}, {Nelles}, {Orru}, {Paas}, {Pandey},
  {Pekal}, {Pizzo}, {Polatidis}, {Reich}, {Rowlinson}, {Schwarz}, {Shulevski},
  {Sluman}, {Smirnov}, {Sobey}, {Soida}, {Thoudam}, {Toribio}, {Vermeulen},
  {van Weeren}, {Wucknitz}, \& {Zarka}}]{Zu2018}
{Zucca}, P., {Morosan}, D.~E., {Rouillard}, A.~P., {et~al.} 2018, \aap, 615,
  A89, \dodoi{10.1051/0004-6361/201732308}

\end{thebibliography}
\bibliographystyle{aasjournal}

\end{document}